\documentclass[]{elsarticle}
\usepackage{algorithmicx}
\usepackage{algorithm}
\usepackage{algpseudocode}
\usepackage[utf8]{inputenc}
\usepackage{amsmath,amsthm,amsfonts,amssymb,amscd}
\usepackage{lineno,hyperref}
\usepackage{float}
\usepackage{multirow,booktabs}
\usepackage{amsmath}
\usepackage{algorithm}
\usepackage{subcaption}
\usepackage{amssymb}
\usepackage{amsmath} 
\usepackage{bm} 
\usepackage{mathtools} 
\usepackage{mathrsfs}
\usepackage{natbib} 
\usepackage{graphicx}
\usepackage[x11names]{xcolor}
\usepackage[utf8]{inputenc}
\usepackage[T1]{fontenc}
\usepackage[margin=3cm]{geometry}
\usepackage[lighttt]{lmodern}
\usepackage{amssymb, amsmath, mathrsfs, xcolor, float, graphicx, subcaption}
\usepackage{booktabs} 
\usepackage{array} 
\usepackage[title]{appendix}
\journal{arXiv}

\begin{document}

\begin{frontmatter}



\title{Optimising Kernel-based Multivariate Statistical Process Control}


\author[inst1]{Zina-Sabrina Duma\corref{cor1}}

\affiliation[inst1]{organization={LUT University},
            addressline={Yliopistonkatu 34}, 
            postcode={ FI-53850}, 
            city={Lappeenranta},
            country={Finland}}

\author[inst1]{Victoria Jorry}
\author[inst1]{Tuomas Sihvonen}
\author[inst1]{Satu-Pia Reinikainen}
\author[inst1]{Lassi Roininen}

\cortext[cor1]{Zina-Sabrina.Duma@lut.fi}


\begin{abstract}
Multivariate Statistical Process Control (MSPC) is a framework for monitoring and diagnosing complex processes by analysing the relationships between multiple process variables simultaneously. Kernel MSPC extends the methodology by leveraging kernel functions to capture non-linear relationships between the data, enhancing the process monitoring capabilities. However, optimising the kernel MSPC parameters, such as the kernel type and kernel parameters, is often done in literature in time-consuming and non-procedural manners such as cross-validation or grid search. In the present paper, we propose optimising the kernel MSPC parameters with Kernel Flows (KF), a recent kernel learning methodology introduced for Gaussian Process Regression (GPR). Apart from the optimisation technique, the novelty of the study resides also in the utilisation of kernel combinations for learning the optimal kernel type, and introduces individual kernel parameters for each variable. The proposed methodology is evaluated with multiple cases from the benchmark Tennessee Eastman Process. The faults are detected for all evaluated cases, including the ones not detected in the original study.
\end{abstract}


\begin{highlights}
\item Kernel MSPC is performant to detect challenging faults, if optimisation is done right. 
\item Kernel MSPC optimisation can be achieved through optimising Kernel PCR.
\item Kernel Flows is appropriate for optimising individual-parameter kernels.
\end{highlights}


\end{frontmatter}

\section{Introduction}
\label{sec:sample1}

Multivariate Statistical Process Control (MSPC) is a statistical framework that simultaneously monitors and controls processes with interrelated variables \cite{zhao2022anomaly}. Unlike univariate methods that focus on individual parameters, MSPC considers variable correlations, providing a comprehensive view of the behavior of the process \cite{kourti2005application}. This allows the detection of subtle shifts and anomalies that might be missed when examining variables in isolation \cite{Mason2011}. 

MSPC is effectively used across industries, particularly in the chemical sector \cite{pollanen2006dynamic}, where it concurrently analyzes multiple process parameters to ensure product quality and efficiency \cite{ge2012multivariate}. In semiconductor manufacturing, MSPC manages fabrication processes in which various variables affect the performance of the final product \cite{chien2013semiconductor}. In addition, MSPC is applied in pharmaceutical production to ensure consistent drug quality by monitoring critical quality attributes during manufacturing \cite{shi2021pharmaceutical}.

While MSPC offers benefits, it has limitations. Traditional techniques often assume linear relationships and normal distribution, which may not apply in all situations \cite{Kenett2023}. MSPC can also struggle to detect small shifts in process mean or variance, especially with high-dimensional data and complex nonlinear interactions \cite{park2020review}. These issues highlight the need for advanced methodologies like Kernel MSPC to enhance its effectiveness in diverse industrial contexts \cite{yu2022sparse}.

Kernel MSPC extends traditional MSPC techniques by addressing the limitations of linear assumptions \cite{kong2024overview}. It uses kernel functions to transform process data into a higher-dimensional feature space, allowing for the detection of complex, nonlinear relationships and subtle variations that linear methods may overlook \cite{Ge2013}. Kernel MSPC has its limitations. Selecting the appropriate kernel function and its parameters is critical, as poor choices can lead to overfitting or underfitting, affecting monitoring performance. The computational complexity in high-dimensional feature spaces can also limit real-time applications with large datasets. Additionally, interpreting results can be challenging due to transformations into higher-dimensional spaces, complicating the identification of variables linked to detected anomalies \cite{grasso2015comparison}.

Optimising kernel parameters in Kernel-based MSPC is essential for effective process monitoring and fault detection. Traditional methods use grid search or cross-validation to evaluate combinations of kernel parameters, aiming to find settings that minimize monitoring errors \cite{jemwa2006kernel}. Advanced optimisation algorithms like Bayesian optimisation and genetic algorithms efficiently explore parameter spaces, utilizing an optimisation function that minimizes the false alarms and maximizes the correct identification of faulty observations \cite{jia2012optimization}. However, these methods have a few limitations or areas of improvement: (i) grid search can become time-inefficient when the optimal value resides in a narrow interval; (ii) the existing optimisation methods consider prior knowledge on the kernel function type - the kernel function cannot be learned in the optimisation; (iii) only one kernel parameter is optimised for the whole dataset, which might not be optimal to detect all possible faults.

Kernel Flows (KF) is a methodology that learns optimal kernel functions directly from data, improving the performance of kernel-based models in capturing complex, nonlinear relationships \cite{OWHADI201922}. By iteratively adjusting kernel parameters, it minimizes loss from reducing interpolation points, enhancing model generalization. This approach has been applied to various machine learning tasks, including regression and classification, resulting in increased accuracy and robustness \cite{DUMA2024105238}. This paper explores optimising Kernel MSPC parameters with KF. 

The novelty of the study comes in multiple aspects: (I) the utilization of KF for optimising the kernel parameters in K-MSPC; (II) learning the kernel function for optimal MSPC, alongside the kernel parameters; (III) fitting individual kernels for each of the process variables, for hard-to-catch fault detection; (IV) utlisation of training parameters to gain insights into variable importance in the model.

The method is illustrated with various case studies of simulated faults from the benchmark fault detection Tennessee Eastman Process dataset \cite{russell2000fault}. The article is structured as follows: Section \ref{ssec:matmet} presents the process data and the base mathematical methods utilized, Section \ref{ssec:resultsanddiscussion} presents the results of the KF-optimised MSPC, and the Section \ref{ssec:conclusion} concludes the findings.

\section{Materials and methods}\label{ssec:matmet}

\subsection{Process and data description}

\textit
In this study, the Tennessee Eastman (TE) process is utilised, with the faults simulated in Russell \textit{et al.}, 2000 \cite{russell2000fault}. On this well-established benchmark process, the proposed methods will be applied to demonstrate their efficiency, and to aid in further method comparisons. The TE process model is a realistic simulation program of a chemical plant widely accepted as a benchmark for control and monitoring studies. The process is described in detail in \cite{downs1993plant}, and has with 5 significant units: reactor, condenser, compressor, separator, and stripper. The process has 2 products from 4 reactants, an inert and a by-product, making a total of 8 components denoted \textbf{A}-\textbf{H}. The process measures 52 variables, which are expanded in \textbf{Appendix A}, of which 41 are process variables and 11 are manipulated variables. 

From the total 21 faults presented in \cite{russell2000fault}, four have been selected for visualisation in the present paper (summarised in Table~\ref{tab:1}) The datasets have a training partition, where the fault occurs throughout the whole process operation, and a testing partition, which introduces the fault after 8 simulation hours. The sampling time is every three minutes. 

Because prior information on the mathematical model of the TE process is not available, the Process Monitoring and Fault Diagnosis (PM-FD) system must be constructed solely based on process data. The data sets in \cite{chiang2000fault} are commonly accepted for PM-FD studies. 

\begin{table}[h]
\centering
\caption{Descriptions of considered faults in TE process. * Faults that were not detected in the original study through conventional methods. \cite{russell2000fault} }\label{tab:1}%
\begin{tabular}{llllll}
\hline\noalign{\smallskip}
Fault & Fault & Variable & Fault & Purpose \\
number&  ID &   & affected & type \\
\noalign{\smallskip}\hline\noalign{\smallskip}
1    & E1   & A/C feed ratio, B composition constant & step & Validation \\
2    &  E2  & B composition, A/C ratio constant & step & Testing \\
3    & D1*   & D feed temperature  & step & Validation \\
9    & D2*   & D feed temperature & random variation & Testing \\
\noalign{\smallskip}\hline
\end{tabular}
\end{table}

\subsection{Mathematical methods}

The present section gathers the mathematical methods utilised for generating the MSPC control charts, both in the traditional and kernel PCA versions. Also, the mathematical principles for the parameter optimisation are detailed. 

\subsubsection{PCA and Kernel PCA}

Principal Component Analysis (PCA) \cite{bro2014principal}, is a matrix decomposition method that is characterised by the equation:
\begin{equation}
    \mathbf{X} = \mathbf{T} \mathbf{P}^\intercal + \mathbf{E}
\end{equation}
where, for a number of principal components $H$, $\mathbf{T}$ is a matrix of scores, $\mathbf{P}$ is a matrix of loadings, and $\mathbf{E}$ is a matrix of residual variation.

In Kernel Principal Component Analysis (K-PCA) \cite{jia2012optimization, fezai2018online}, the principal component decomposition is occurring in a Reproducing Kernel Hilbert Space, RKHS ($\mathscr{H}$). For $n$ data points in $d$ original dimensions, $\mathbf{x}_i \in R^d$, where $i = 1, ..., n$, there is a feature map to $\mathscr{H}$  ($\phi: \mathbf{x}_i \to \phi(\mathbf{x}_i) $), which maps the original datapoints as functions in the RKHS. In this space the kernel function is defined as the inner product between feature maps 
$ \textbf{K} =\langle\phi(\mathbf{x}), \phi(\mathbf{y})\rangle $. 


The covariance matrix ($\mathbf{C}$) in $\mathscr{H}$ can be expressed as
\begin{equation}
    \mathbf{C} = \frac{1}{n} \sum_{i=1}^n \sum_{j=1}^n \phi(\mathbf{x}_j) \phi(\mathbf{x}_i)^\intercal.
    \label{eq:cov}
\end{equation}

The eigenvalue problem can be solved as
\begin{equation}
    \boldsymbol{\lambda} v = \mathbf{C}, 
    \label{eq:v1}
\end{equation}
where $\boldsymbol{\lambda}$ is positive and represents the eigenvalues and $v$ represents the solutions, which should be in the span of $\phi(\mathbf{x}_1), ..., \phi(\mathbf{x}_n) $. So, we can express it as
\begin{equation}
    v = \sum_{i=1}^n \alpha_i \phi(\mathbf{x}_i).
    \label{eq:v2}
\end{equation}

According to Eq.\eqref{eq:v2}, Eq.\eqref{eq:v1} can be written as
\begin{equation}
    n \boldsymbol{\lambda} \alpha = \mathbf{K} \alpha,
\end{equation}
where $K_{i,j} = \langle\phi(\mathbf{x}_i), \phi(\mathbf{x}_j)\rangle$ is an element of the kernel matrix $\mathbf{K}$. 

The principal components scores $t$ for the $h$-th principal component are extracted by projecting $\phi(\mathbf{X})$ onto the $h$-th eigenvector in $\mathscr{H}$
\begin{equation}
    \mathbf{t}_h = \sum_{i=1}^n \alpha_i^h \langle\phi(\mathbf{x}_i), \phi(\mathbf{X})\rangle.
\end{equation}

To map the data from the original data space to $\mathscr{H}$, a kernel function can be utilised. A widely utilised kernel is the Mat{\'e}rn kernel, that can be expressed in in the general form as:

\begin{equation}
    \mathsf{k}(\mathbf{x}, \mathbf{y}; \boldsymbol{\theta})_M =  \gamma^2 \frac{2^{1-\nu}}{\Gamma(\nu)} \left( \sqrt{2\nu}\frac{\Vert\mathbf{x} - \mathbf{y}\Vert}{\sigma} \right)^{\nu} K_{\nu} \left( \sqrt{2\nu}\frac{\Vert\mathbf{x} - \mathbf{y}\Vert}{\sigma} \right) ,
\end{equation}
where $\boldsymbol{\theta}$ is the hyperparameter vector and includes $\{ \gamma^2, \sigma, \nu \}$, $K_{\nu}$ is the modified Bessel function of the second kind and $\Gamma()$ is the Gamma function.  When $\nu$ is $\{ \frac{1}{2}, \frac{3}{2}, \frac{5}{2}, ...\}$, the Mat{\'e}rn kernel takes simpler forms that are easier to evaluate. If $\nu$ has a smaller value (i.e. $\frac{1}{2}$), the kernel is very rough. As $\nu$ approaches $\infty$, the smoother Gaussian kernel is obtained

\begin{equation}
    \mathsf{k}(\mathbf{x}, \mathbf{y}; \boldsymbol{\theta})_G = \gamma^2 \exp \left( - \frac{\Vert\mathbf{x} - \mathbf{y}\Vert^2}{2\sigma^2}\right) \text{.}
\end{equation}

If data has outliers, a Cauchy kernel can be utilised:
\begin{equation}
    \mathsf{k}(\mathbf{x}, \mathbf{y}; \boldsymbol{\theta})_{C} = \gamma^2 \frac{1}{1 + \frac{\Vert \mathbf{x} - \mathbf{y} \Vert^2 }{\sigma^2}} \text{.}
    \label{eq:cauchy}
\end{equation}

If the optimal kernel function for a dataset is not known, a combination of kernels can be utilised \cite{duma2024kernel}. Individual kernel parameters can be fitted for each of the process variables:

\begin{equation}\label{eq:addKernels}
    \mathsf{k}(\textbf{X},\textbf{X}; \boldsymbol{\theta}) =\mathsf{k}(\mathbf{x}_1, \mathbf{x}_1; \boldsymbol{\theta}_1) + \mathsf{k}(\mathbf{x}_2, \mathbf{x}_2; \boldsymbol{\theta}_2) + ... + \mathsf{k}(\mathbf{x}_d, \mathbf{x}_d; \boldsymbol{\theta}_d),
\end{equation}
where $d$ is the number of original dimensions and $\boldsymbol{\theta} = [\boldsymbol{\theta}_1, \boldsymbol{\theta}_2, ..., \boldsymbol{\theta}_d]^{\intercal}$. 

The centering of the kernel matrix in $\mathscr{H}$ is expressed with
\begin{equation}
    \mathsf{\widetilde{k}}(\mathbf{x}, \mathbf{y}; \boldsymbol{\theta}) = \left( \mathbf{I} - \frac{1}{n} \mathbf{1}_n \mathbf{1}_n^{\intercal} \right) \mathsf{k}(\mathbf{x}, \mathbf{y}; \boldsymbol{\theta})\left( \mathbf{I} - \frac{1}{n} \mathbf{1}_n \mathbf{1}_n^{\intercal} \right),
\end{equation}
where $\mathbf{1}_n$ is a vector with the size of the number of data points $n$. Each of the elements of the $\mathbf{1}_n$ vector is divided by the number of data points $\left( \frac{1}{n} \right)$. 

\subsubsection{MSPC and Kernel MSPC}

Multivariate Statistical Process Control charts include charts that tackle the modelled variation, \textit{Hotelling's} T$^2$ charts, and charts that capture new variation from the modelled one, the Squarred Prediction Error (SPE) charts. To emphasize the prediction error is for the reconstruction of the input matrix from the model, the SPE charts are often denoted as $\mathrm{SPE_x}$. Considered together, the $\mathrm{SPE_x}$ and T$^2$ control charts are complementary and most efficiently utilised together. 

Hotelling's T$^2$ statistic is utilised to identify model outliers - measurements that have scores far from the model centre \cite{bro2014principal}. For the $i$-th sample, it can be computed as follows:
\begin{equation}
    \mathrm{T}^2_i = \frac{\mathbf{t}_i^\intercal (\mathbf{T}^\intercal\mathbf{T})^{-1}\mathbf{t}_i}{n - 1}
\end{equation}
where $\mathbf{t}_i$ is a vector containing the scores for the $i$-th sample, $\mathbf{T}$ is the complete score matrix, and $n$ is the total number of samples.

If the unmodelled variation is stored in the residual matrix $\mathbf{E} = \mathbf{X} - \mathbf{T} \mathbf{P}^\intercal $, the SPEx for a sample $i$ can be expressed as:
\begin{equation}
    \mathrm{SPEx}_i = \sum_{j=1}^{m} e_{i,j}^2
\end{equation}
where $e_{i,j}$ is an element of the residual matrix for the $i$-th sample and $j$-th variable.

For K-PCA, Hotelling's T$^2$ can be computed in a similar fashion. The squared prediction error ($\mathrm{SPE_x}$) \cite{kallas2017fault} is as follows:
\begin{equation}
    \mathrm{SPEx} = \mathsf{k}(\mathbf{X}, \mathbf{X}) - \mathsf{k}(\mathbf{X})^\intercal \mathbf{C} \mathsf{k}(\mathbf{X}),
\end{equation}
where $\mathsf{k}(\mathbf{X}) = [\mathsf{k}(\mathbf{x}_1, \mathbf{X}), ..., \mathsf{k}(\mathbf{x}_n, \mathbf{X})]^\intercal$ and $\mathbf{C}$ is the covariance matrix expressed in Eq.~\eqref{eq:cov}.

The confidence limits for the T$^2$ index can be computed as a function of the $F$-distribution
\begin{equation}
    \mathrm{T}^2_\mathrm{lim, \alpha} = \frac{h(n-1)}{n-h}F_{h,n-h,\alpha},
\end{equation}
where $n$ is the number of samples and $h$ is the number of PCs. Another way to compute the indicators is by assuming a Gaussian distribution, where the warning limit (95\%C.I.) is computed as
\begin{equation}
    \mathrm{T}^2_\mathrm{warning} = \mu_{\mathrm{T}^2} + 2\sigma_{\mathrm{T}^2}\text{,}
    \label{eq:19}
\end{equation}
and the alarm limit (99\%C.I.) is computed as
\begin{equation}
    \mathrm{T}^2_\mathrm{alarm} = \mu_{\mathrm{T^2}} + 3\sigma_{\mathrm{T^2}} \text{.}
    \label{eq:20}
\end{equation}

The confidence limits for the $\mathrm{SPE_x}$ are expressed as a function of the $\chi^2$-distribution
\begin{equation}
    \mathrm{SPE_x}_\mathrm{lim} = \frac{\sigma^2_{\mathrm{SPE_x}}}{2\mu_{\mathrm{SPE_x}}} \chi^2_l
    \label{eq:21}
\end{equation}
where $l = \frac{2\mu^2_{\mathrm{SPE_x}}}{\sigma^2_{\mathrm{SPE_x}}}$, $\mu_{\mathrm{SPE_x}}$ is the $\mathrm{SPE_x}$ mean and $\sigma^2_{\mathrm{SPE_x}}$ is here the $\mathrm{SPE_x}$ variance.



\subsubsection{Kernel Flows}

Kernel Flows (KF) is a kernel parameter learning technique proposed by Owhadi \& Yoo, 2019 \cite{OWHADI201922}, that utilises stochastic gradient descent to minimise a loss function obtained in a cross-validative manner. The loss function for each iteration is based on the difference of the model calibrated with all data points and the model calibrated with half of the data points. If the kernel is correctly chosen, the norm in the reproducing kernel Hilbert space ($\mathscr{H}$) should be similar for both the full-points model and the reduced-points model.

The proposed method has similarities with the Kernel Flows approach, including its iterative and cross-validated learning, and parameter updating with stochastic gradient descent, and to the modified version in Lamminpää \textit{et al.}, 2024 \cite{lamminpaa2024forward} that has a Leave-One-Out (LOO) approach and computes an $\ell^2$-norm loss. The method is summarised in Algorithm \ref{alg:KPCR}. 

Compared to the original approaches, the proposed one aims to produce process control limits by solving a classification problem that differentiates between the normal and the faulty states of a process. 
The loss function for a sub-iteration $s$ is
\begin{equation}\label{eq:loss1}
    \rho_s = 1 - \frac{\eta_n + \eta_f}{2}
\end{equation}
where $\eta_n$ is the proportion of normal-functioning samples that are classified as normal ("class 0") and $\eta_f$ is the proportion of faulty operation samples that are classified as faulty ("class 1"). Thus, the kernel parameters are optimised in a way that facilitates a good separation between the normal and faulty states of the process.

\begin{algorithm}[H] 
    \caption{Kernel parameter optimisation workflow} \label{alg:KPCR}
    \textbf{\textit{Input:}} Initial kernel parameters ($\theta_0$), normal-functioning process data ($\mathbf{X}^{normal}$), faulty-functioning process data ($\mathbf{X}^{faulty}$), number of PCs ($H$), number of iterations ($I$), number of samples extracted in a sub-interation ($ns$), number of sub-iterations ($S$), learning rate ($\alpha$) \textbf{\textit{Output:}} optimised kernel parameters ($\theta_I$).
    \begin{algorithmic}[1]
    \For{$i = 1$ to $I$}
        \For{$s = 1$ to $S$}
            \State $\mathbf{X}^{normal}_{s} \gets  \mathbf{X}^{normal}_{\pi, \cdot}$ where $\pi$ is an $ns$ sample extraction from a random permutation.
            \State $\mathbf{X}^{faulty}_{s} \gets  \mathbf{X}^{faulty}_{\pi, \cdot}$
            \State $\mathbf{X}_{s} \gets \left[ \mathbf{X}^{faulty}_{s};  \mathbf{X}^{normal}_{s}\right]$ 
            \State $ \mathbf{K} \gets k(\mathbf{X}_{s},\mathbf{X}_{s}; \boldsymbol{\theta}_i)$
            \State $\widetilde{\mathbf{K}} \gets (\mathbf{I} - \frac{1}{N}\mathbf{1}_N\mathbf{1}_N^\intercal)\mathbf{K} (\mathbf{I} - \frac{1}{N}\mathbf{1}_N\mathbf{1}_N^\intercal)$
            \State $ \mathbf{T} \gets \mathrm{kpca}(\widetilde{\mathbf{K}}, H) $
            \State $\mathbf{y} \gets \left[ \bold{1}_{ns}; \bold{0}_{ns} \right]$
            \State $ \mathbf{b} \gets \mathbf{T}^{\dagger} \mathbf{y}$
            \State $ \rho_s \gets 1 - \frac{\eta_n + \eta_f}{2}$ 
        \EndFor
        \State $\bar{\rho}_i \gets \sum_{s=1}^S \rho_s$
        \State $\nabla_{\theta, i} \gets \mathrm{diff}(\theta_i, \bar{\rho}_i)$
        \State $\theta_{i+1} \gets \alpha \nabla_{\pmb \theta, i} f(\theta_i)$
    \EndFor
    \end{algorithmic}   
\end{algorithm}

The results will be compared to grid search, and other benchmark optimisation techniques such as the Genetic Algorithm (GA) and Nedler-Mead simplex. The success of the convergence parameters both in the proposed method and the benchmark techniques is evaluated with a similar cost function,  based on the correct monitoring rate (CMR) \cite{jia2012optimization}, where
\begin{equation}\label{eq:loss2}
    \rho_{\eta} = 1 - \frac{\eta_n^{eval} + \eta_f^{eval}}{2} 
\end{equation}
$\eta_n^{eval}$ is the proportion of normal-functioning samples that are under the alarm limit in MSPC, $\eta_f^{eval}$ is the proportion of faulty operation samples that are above the alarm limit in MSPC. The main difference between the cost functions in the optimiser Eq.~\eqref{eq:loss1} and the final evaluation Eq.~\eqref{eq:loss2} is that, in the optimiser, the classification rate is monitored, and in the latter case, the correct monitoring rate in MSPC chart is followed.

\subsection{Proposed procedure}

\begin{figure}[H]
    \centering
    \includegraphics[width=0.8\linewidth]{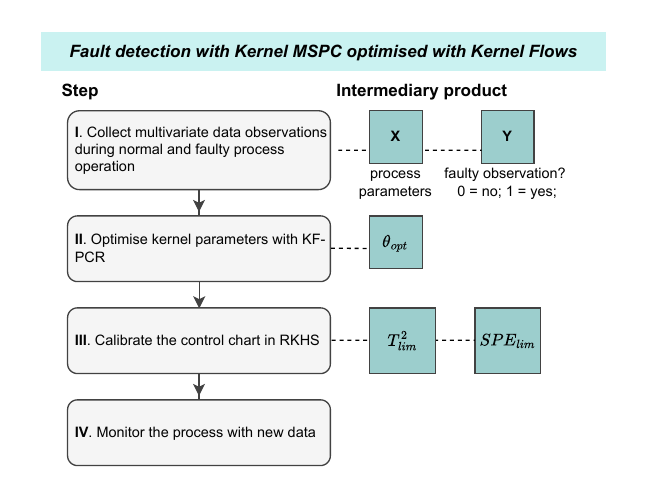}
    \caption{Proposed fault detection workflow.}
    \label{fig:workflow}
\end{figure}

The diagram in Fig. \ref{fig:workflow} summarises the proposed procedure based on the mathematical methods presented. The process data is collected, both in normal-functioning and sub-optimal or faulty operations. Simulations of the process operation in possible faulty conditions can be utilised. 

After there exists data on both the normal and faulty process, the kernel parameters are learned with the Kernel Flows-inspired procedure presented in the current study. The control charts are created with the K-PCA model with optimised parameters. 

When new data arrives, it is projected onto the K-PCA model, and further into the control charts. The position of the projections in relation to the control chart limits are evaluated, and the state of the process is determined (in-control or faulty). Further investigation on the causes of the fault can be conducted through contribution analysis, but the issue is not the subject of the present paper.

In our examples, the dataset has three parts: (a) a healthy-functioning monitoring, (b) a faulty-functioning calibration set and (c) a test partition, where the process starts normal-functioning and the fault develops after eight hours. The first two parts (a, b) will be utilised for model calibration and kernel parameter optimisation (Steps I - III in \ref{fig:workflow}), and the test partitions (c) will be utilised to evaluate and exemplify a monitoring case (Step IV). 

\section{Results and discussion}\label{ssec:resultsanddiscussion}

\subsection{Regular MSPC results}

\begin{figure}[H]
    \centering
    \includegraphics[width=0.56\linewidth]{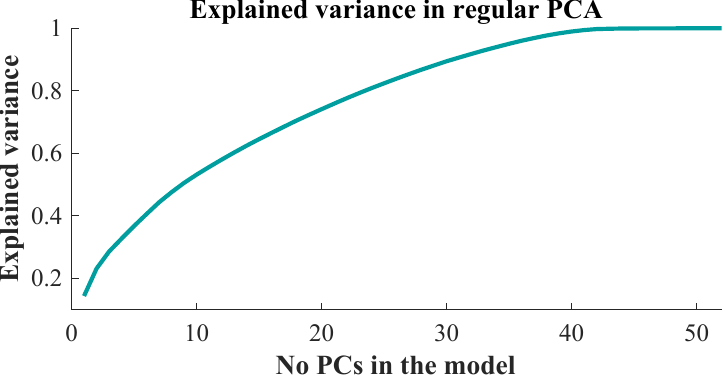}
    \caption{Explained variance in the regular PCA model of the normal-functioning data.}
    \label{fig:explainedVariancePCA}
\end{figure}

Before going into the more complex kernel MSPC, the linear MSPC results are discussed. In the regular PCA, if we consider the explained variance of the PCs (Fig.~\ref{fig:explainedVariancePCA}), we can observe that there are no dominant variational profiles in the dataset. The cumulative explained variance of the first two PCs sums up to about 20\% of the variation, and the increase in variance explained after 4 PCs is very shallow. The increase in model complexity does not justify adding more PCs. This is the first clue that suggests a non-linear approach might be optimal for the dataset.  

\begin{figure}[H]
    \centering
    \begin{subfigure}[b]{0.59\linewidth}
        \centering
        \includegraphics[width=\linewidth]{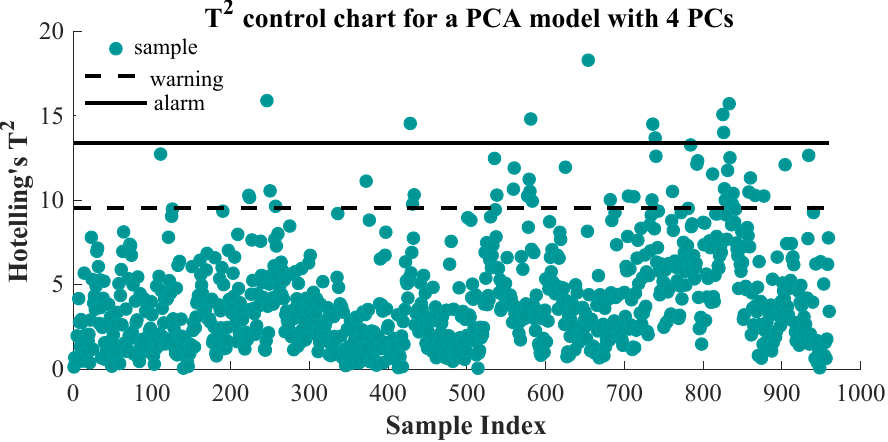}
        \caption{}
        \label{fig:healthyPCAT2}
    \end{subfigure} 
    \hfill
    \begin{subfigure}[b]{0.56\linewidth}
        \centering
        \includegraphics[width=\linewidth]{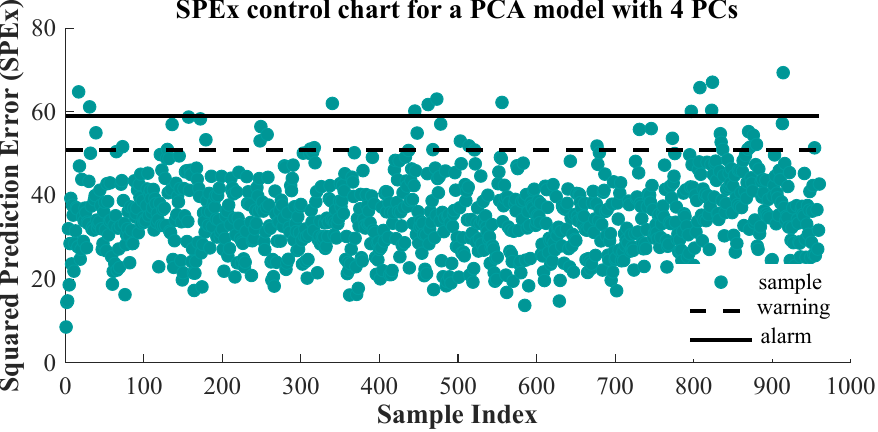}
        \caption{}
        \label{fig:SPExHealthyPCA}
    \end{subfigure}
    \caption{The T$^2$ and $\mathrm{SPE_x}$ control chart calibration with regular PCA and normal-functioning process data.}
    \label{fig:regularPCA}
\end{figure}

In Figure~\ref{fig:healthyPCAT2}, one can observe the calibration of the T$^2$ control chart with samples from a regular process functioning. The points represent Hotelling's T$^2$ scores, which express how well a data point fits in the model, in this case in a 4 PCs PCA model. The closer a data point is to the model's origin, the lower the T$^2$ score is. It is, thus, a measure of consistency with the variation included in the model. Warning (95\% CI limits) and alarm (99\% CI limits) are calculated as in Eq.\eqref{eq:19} and Eq.\eqref{eq:20} . It is expected that 5\% normal functioning samples are found outside the control chart limits for the 95\% C.I., and respectively 1\% for the 99\% C.I. These are denoted "false alarms", and their minimisation is desired. 

The $\mathrm{SPE_x}$ control charts represent the variation that is not included in the PCA model, the one that is not included in the model. When a process goes out of control, the $\mathrm{SPE_x}$ control chart usually signals it first. The control chart can be visualised on Fig~\ref{fig:SPExHealthyPCA}. The limits are computed as in Eq.\eqref{eq:21}. 

\begin{figure}[H]
    \centering
    \begin{subfigure}[b]{0.47\linewidth}
        \centering
        \includegraphics[width=\linewidth]{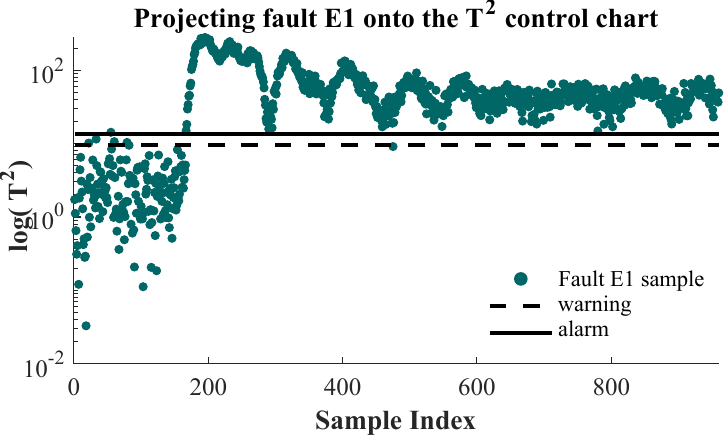}
        \caption{E1 $log(T^2)$}
        \label{fig:PCA_E1_t2}
    \end{subfigure}
    \hfill
    \begin{subfigure}[b]{0.49\linewidth}
        \centering
        \includegraphics[width=\linewidth]{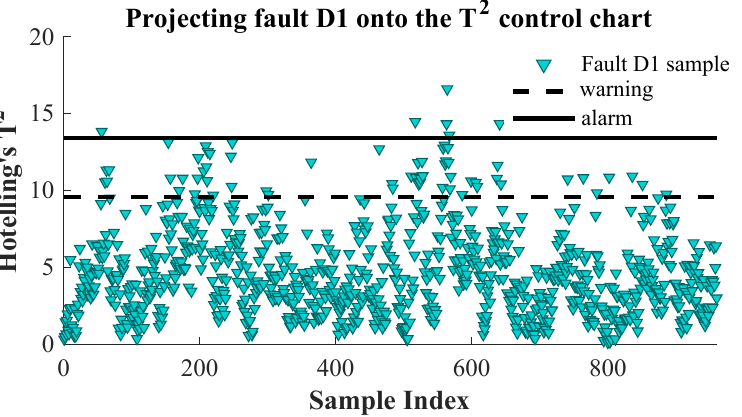}
        \caption{D1 T$^2$}
        \label{fig:PCA_D1_t2}
    \end{subfigure}
    \hfill
    \begin{subfigure}[b]{0.49\linewidth}
        \centering
        \includegraphics[width=\linewidth]{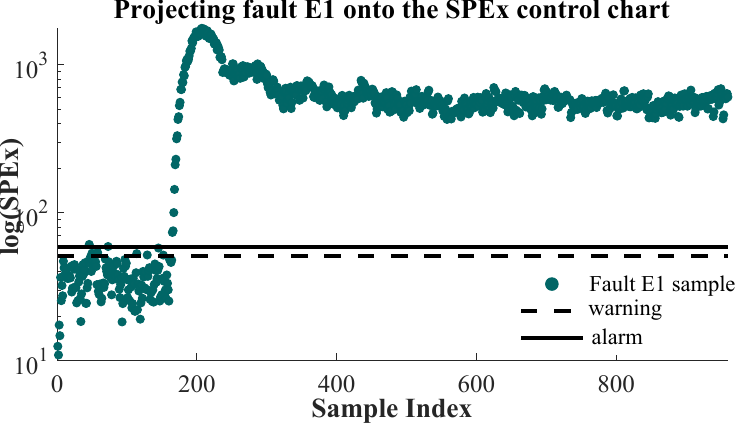}
        \caption{E1 $log(\mathrm{SPE_x})$}
        \label{fig:PCA_E1_SPEx}
    \end{subfigure}
    \hfill
    \begin{subfigure}[b]{0.49\linewidth}
        \centering
        \includegraphics[width=\linewidth]{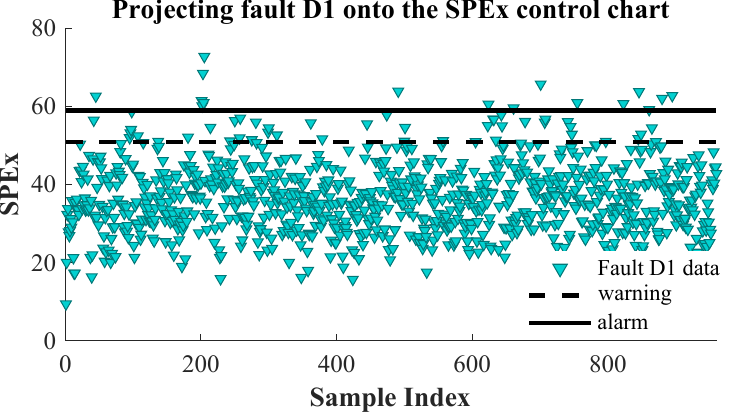}
        \caption{D1 $\mathrm{SPE_x}$}
        \label{fig:PCA_D1_SPEx}
    \end{subfigure}
    \begin{subfigure}[b]{0.69\linewidth}
        \centering
        \includegraphics[width=0.95\linewidth]{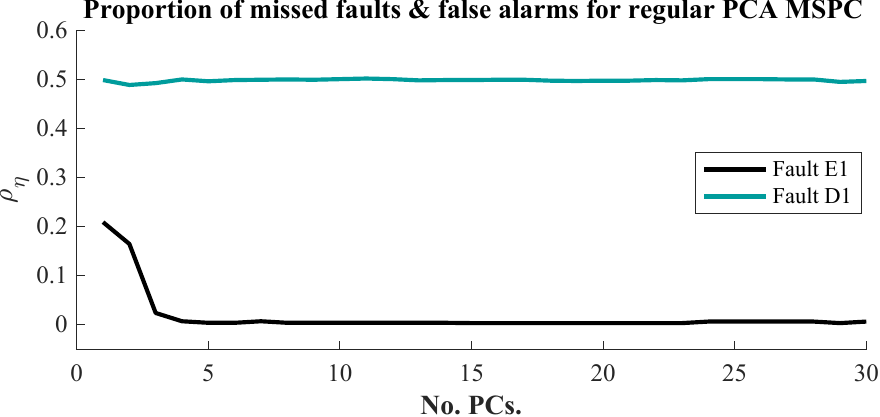}
        \caption{}
        \label{fig:lossRegularPCA}
    \end{subfigure}
    \hfill
    \caption{Capturing the faults using the regular PCA control charts. While the "easy" E1 fault is captured in both T$^2$ (a) $\mathrm{SPE_x}$ (d) control charts, the "difficult" fault is not captured in any of them. Increasing the number of PCs in the model (e) is not helpful in fault identification.}
    \label{fig:regularPCAfaultsT2}
\end{figure}

Figures~\ref{fig:PCA_E1_t2} and \ref{fig:PCA_E1_SPEx} present scenarios where the control charts successfully identify the fault both in the included variance and the new variation presented by the faulty data. This fault (\textbf{E1}) can be successfully identified using traditional MSPC charts because it presents deviations from the modeled profiles.  However, the more difficult fault (\textbf{D1}) is not successfully identified by the control charts. As seen in Figures \ref{fig:PCA_D1_t2} and \ref{fig:PCA_D1_SPEx}, the fault is hardly identified. Increasing the number of PCs in the model does not drive an enhancement in the control charts either. Figure~\ref{fig:lossRegularPCA} showcases the identification of the fault in the control chart. A constant 0.5 $\rho$ value shows that regardless of the PCs included in the model, not only the normal-functioning 8h sampling window is in-control lines, but also the fault that develops after 8h. Thus, a linear approach is not appropriate for identifying the \textbf{D1} fault. 

\subsection{Optimising K-MSPC with line search}

In some cases where the linear methods are insufficient for capturing the challenging cases, or the model is more appropriately represented by a non-linear method, the kernelised MSPC is a good solution. However, the parameter optimisation for Kernel MSPC is non-procedural, and it often implies a grid search for optimising the kernel parameter. 

\begin{figure}[H]
    \centering
    \begin{subfigure}[b]{0.69\linewidth}
        \centering
        \includegraphics[width=0.9\linewidth]{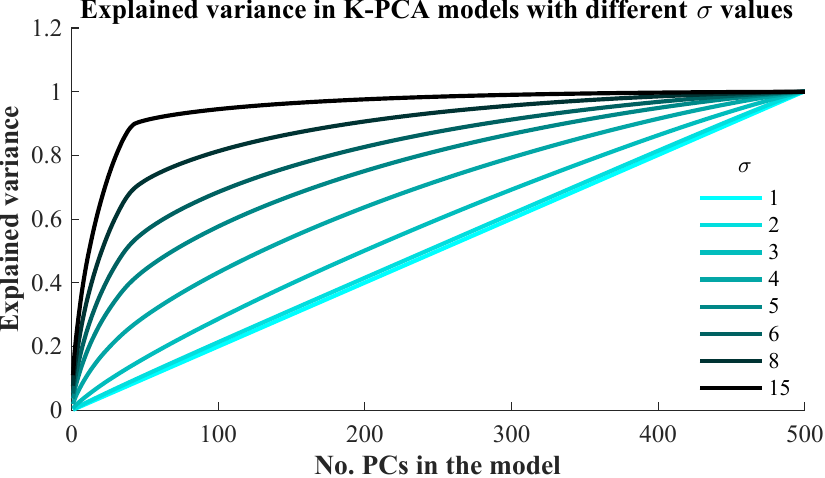}
        \caption{}
        \label{fig:kpcaexpl}
    \end{subfigure}
    \hfill
    \begin{subfigure}[b]{0.69\linewidth}
        \centering
        \includegraphics[width=0.95\linewidth]{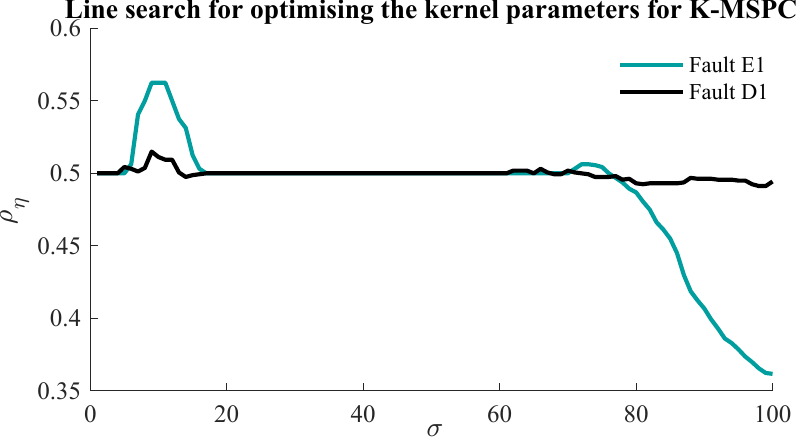}
        \caption{}
        \label{fig:kpcaloss}
    \end{subfigure}
    \caption{(a) The variation explained by PCs in K-PCA is dependent on the shape of the kernel. In the showcased methods, the kernel function utilised is a Gaussian one. (b) The loss value for line-search of the Gaussian kernel parameter, for both E1 and D1 faults.}
    \label{fig:kPCAexplained}
\end{figure}

In K-PCA, the explained variance of the PCs is dependent on the kernel parameter used. The training kernel matrix is a symmetric matrix that has as many rows and columns as the initial number of observations. This drives the maximum number of possible PCs to increase if the number of rows is larger than the number of variables, which is also the case here. In Fig.~\ref{fig:kpcaexpl}, the explained cumulative variation is shown. With an increase in the Gaussian kernel width, the explained variance by the first PCs increases.

When evaluating the kernel parameters through line-search, the easy-to-detect \textbf{E1} fault finds a convergence at a minimum. However, the more difficult fault oscillates around the same $\rho$ value that the linear case resulted. This can indicate that the evaluated range is not sufficient, the discretisation of the line search is too coarse, or fitting one kernel parameter for the whole dataset is not appropriate for the task. Thus, a more specialised technique that is able to select the kernel function and kernel parameters individually for each variable is needed.

\subsection{Kernel MSPC optimised with Kernel Flows}\label{ssec:KFMSPC}

In the literature evaluated on kernel-MSPC, only one kernel parameter is fitted for the whole dataset $\mathsf{k}(\mathbf{X}, \mathbf{X}, \boldsymbol{\theta})$. This scenario works well for the "easy to detect" faults - the ones that work well with the traditional MSPC. This is the case also for the \textbf{E1} fault. 

\begin{figure}[H]
    \centering
    \begin{subfigure}[b]{0.69\linewidth}
        \centering
        \includegraphics[width=\linewidth]{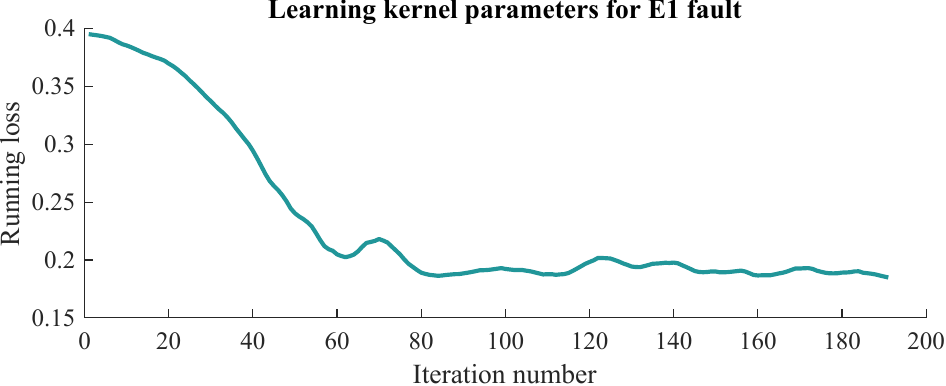}
        \caption{}
        \label{fig:lossE1KPCR}
    \end{subfigure}
    \hfill
    \begin{subfigure}[b]{0.69\linewidth}
        \centering
        \includegraphics[width=\linewidth]{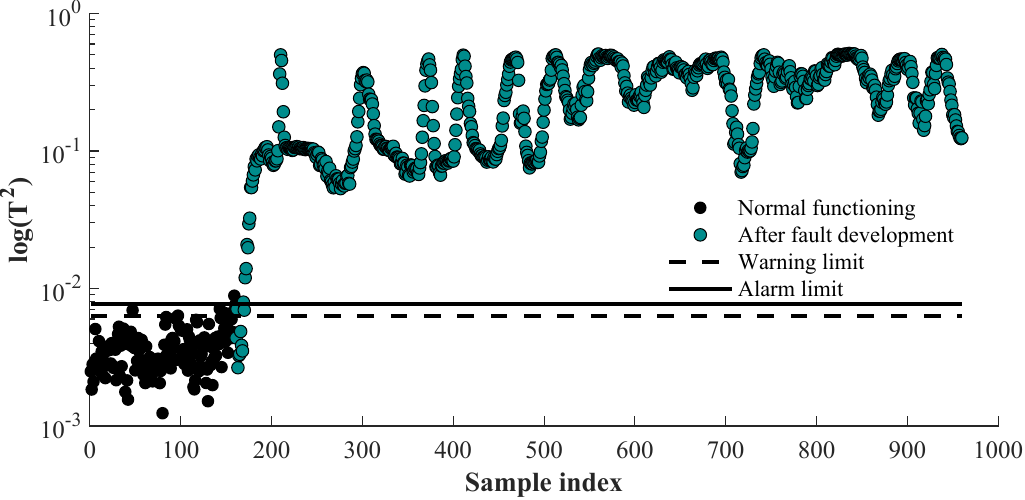}
        \caption{}
        \label{fig:t2kpcrE1}
    \end{subfigure}    
    \caption{(a) Single-kernel parmaeter learning for a Gaussian kernel, trained with the E1 fault and (b) the resulting T$^2$ control chart with optimised kernel parameters.}
    \label{fig:E1KPCRloss}
\end{figure}

Figure~\ref{fig:lossE1KPCR} showcases the KF-PCR learning loss, that is, minimising the percentage of misclassifications between healthy and faulty observations. It is noted that the loss converges to a minimum in relatively few iterations and is stabilised there. The loss is expected to approach a minimum, but not necessarily 0, since the model optimises a classification with a discriminant analysis - and not the control charts directly. As it can be seen in Fig.~\ref{fig:t2kpcrE1}, for the test case where the process starts in control and develops a fault after 8h, the control chart correctly identifies the fault. 

However, in the more difficult cases, such as the \textbf{D1} case, fitting one kernel parameter for the whole dataset is not optimal. The fault \textbf{D1} cannot be identified by regular MSPC nor with single-kernel parameter optimisation techniques. 

Another option would be to fit individual kernel parameters for each of the variables and sum the individual kernel matrices together. However, for this case study, doing so for 52 variables increases the complexity of the grid search, and the task becomes computationally expensive and time-consuming. 

\begin{figure}[H]
    \centering
    \begin{subfigure}[b]{0.69\linewidth}
        \includegraphics[width=\linewidth]{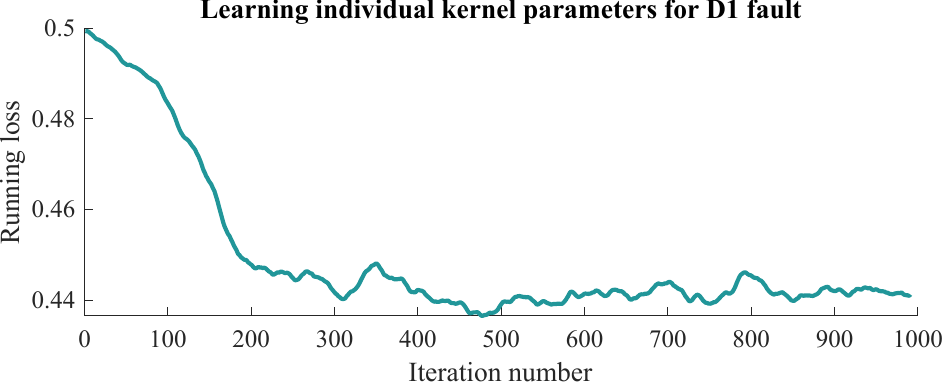}
        \caption{}
        \label{fig:trainingProcess}
    \end{subfigure}
    \hfill
    \begin{subfigure}[b]{0.69\linewidth}
        \includegraphics[width=\linewidth]{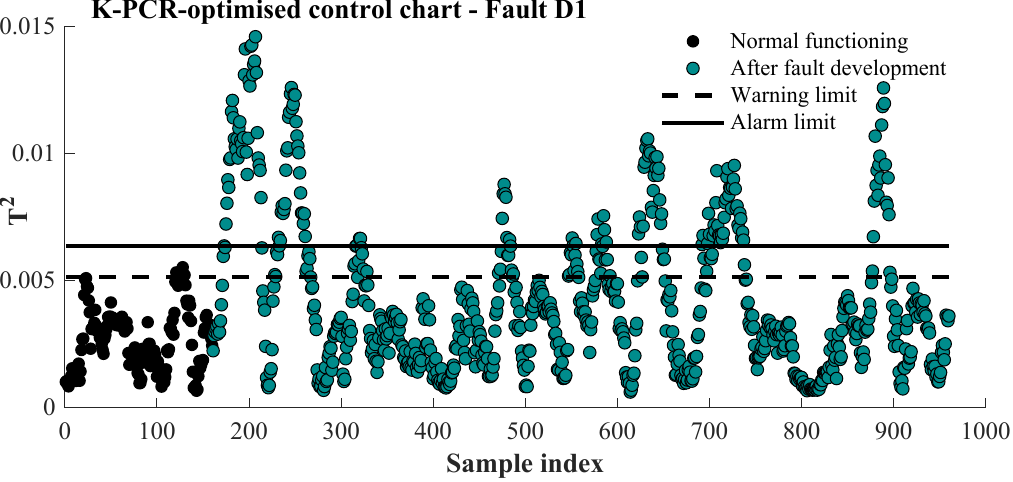}
        \caption{}
        \label{fig:controlCharts}
    \end{subfigure}
    \caption{(a) Kernel flow loss convergence for the D1 K-PCR optimisation with 4 PCs and a Cauchy kernel, and (b) the resulting control chart for the converged parameters.}
    \label{fig:D1controlChart}
\end{figure}

Optimising K-PCR with Kernel Flows to discriminate between the healthy and faulty states by fitting individual kernel parameters to each variable is a solution to the issue. As seen in Fig.~\ref{fig:trainingProcess}, the loss converges to a minimum in less than 300 iterations, and, for the first time, the \textbf{D1} fault is captured in a control chart (Fig.~\ref{fig:controlCharts}). 

When fitting individual kernel parameters to each of the variables, not only do we obtain a better fault identification, but we also gain interpretability on the model. By evaluating the convergence of the kernel parameters, we can quantify the amount of variation that is included from each variable. As the Cauchy kernel is defined in Eq.~\eqref{eq:cauchy}, the smaller the kernel parameter, the less variation from that variable is included.

In this case, the fitted kernel parameters show small fitted parameters for \textbf{x02} and \textbf{x03} variables, which represent the feed values of \textbf{A} and \textbf{D} compounds. A small amount of variation from these parameters will be included. However, variables \textbf{x20} (the stripper stream flow) and \textbf{x49} (the separator liquid flow) have a lot of variation included, as their convergence parameters are large. Since the fault is occurring in a metric that is not captured within the process variables (the \textbf{D} feed temperature), the effects seen in the process are indirect. 

\begin{figure}[H]
    \centering
    \begin{subfigure}[b]{0.69\linewidth}
        \centering
        \includegraphics[width=\linewidth]{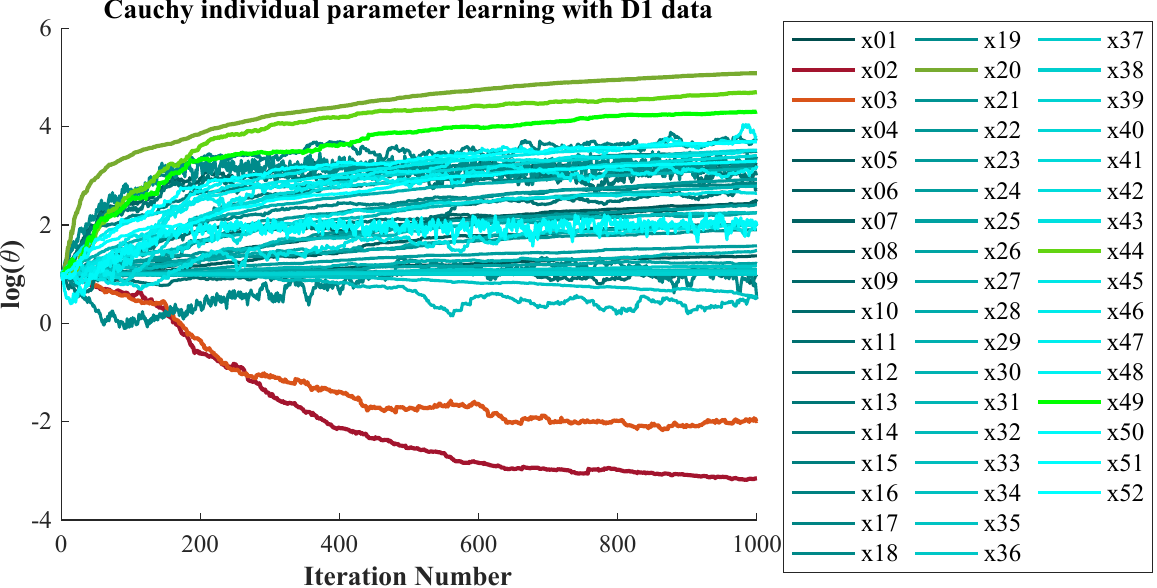}
        \caption{}
        \label{fig:regularParameterLearningD1}
    \end{subfigure}
    \hfill
    \begin{subfigure}[b]{0.69\linewidth}
        \centering
        \includegraphics[width=\linewidth]{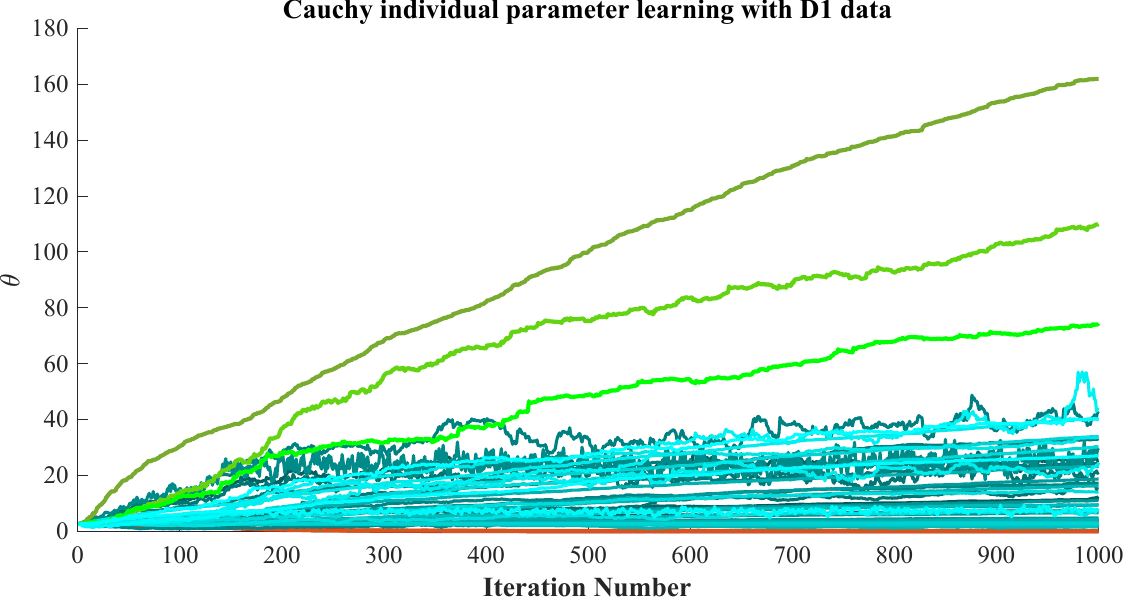}
        \caption{}
        \label{fig:expParameterLearningD1}
    \end{subfigure}
    \caption{The convergence of individual kernel parameters for the Cauchy kernel in the case of \textbf{D1} case learning, where $\theta = \{ \sigma \}$ and $\gamma^2 = 1$ is kept constant. The case where $\boldsymbol{\theta} = \{\sigma, \gamma^2 \}$ is optimised is found in \textbf{Appendix C}.}
    \label{fig:diffCaseD1}
\end{figure}

The previous scenarios consider the parameter tuning on the calibration partion that represents process data having the same fault as in the testing. When calibrating the control charts, it is not always possible to capture all possible faults trough experiment or simulation. Ideally, if a few faults have been simulated and considered, the model can catch the faults regardless if it has seen similar fault data in the calibration phase. 

\begin{figure}[H]
    \centering
    \begin{subfigure}[b]{0.69\linewidth}
        \includegraphics[width=\linewidth]{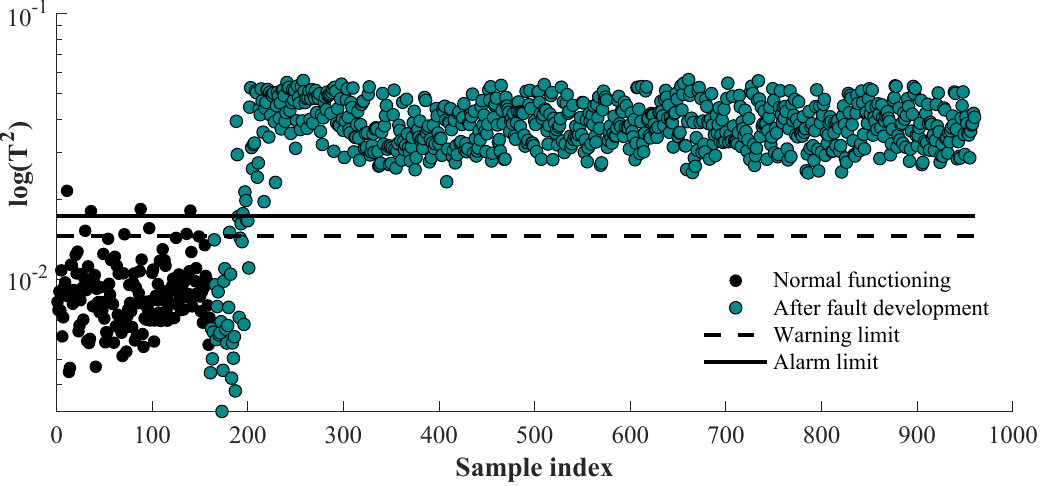}
        \caption{\textbf{E2} fault.}
        \label{fig:e2faultT2}
    \end{subfigure}
    \begin{subfigure}[b]{0.69\linewidth}
        \includegraphics[width=\linewidth]{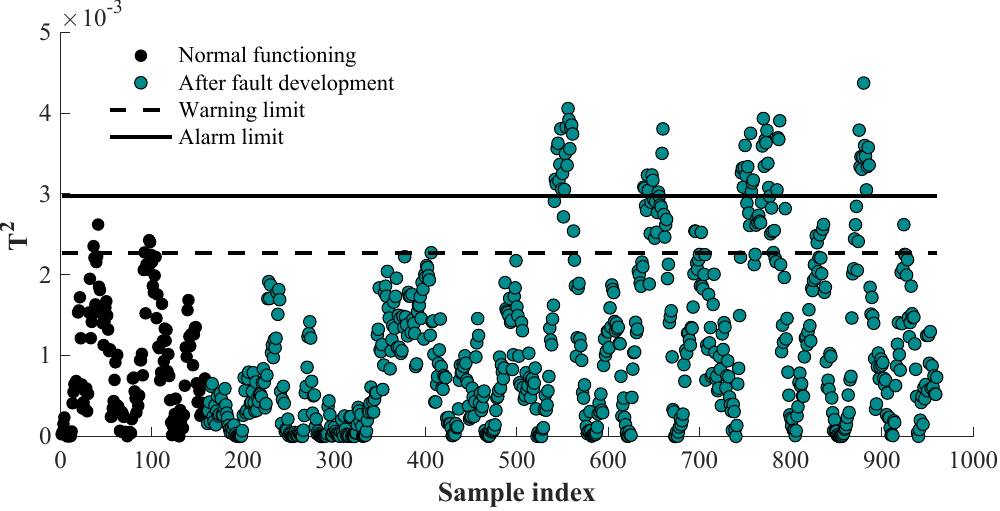}
        \caption{\textbf{D2} fault.}
        \label{fig:d2faultT2}
    \end{subfigure}
    \caption{Parameters optimised with \textbf{E1}+\textbf{D1} faults, and new data projected from \textbf{E2} (a) and \textbf{D2} faults.}
    \label{fig:testingDatasetsProjected}
\end{figure}

Figure \ref{fig:testingDatasetsProjected} presents the control charts resulted when the kernel parameters have been learned with \textbf{E1}+\textbf{D1} faulty process data, and the models have been applied to \textbf{E2} and \textbf{D2} process faults. The resulting control charts illustrate that the faults in the testing datasets \textbf{E2} and \textbf{D2} can still be captured. 

The results of the fault detection of fault \textbf{D2} do not seem impressive when compared to the easier-to-detect \textbf{E1} fault. However, compared to other optimisation methods, such as the Genetic Algorithm \cite{jia2012optimization} and the Nedler-Mead Simplex optimisation utilising the same loss function, the KF-based optimisation method performs better. The results of these methods are presented in \textbf{Appendix D}.

\subsection{Further discussion on advantages and limitations}

The Kernel-MSPC optimisation method presented in this paper has the following advantages when compared to the existing literature:
\begin{itemize}
    \item I. \textit{Difficult faults are captured by individual-parameter models.} It was illustrated in the difficult to catch faults (\textbf{D1}, \textbf{D2}), fitting the same kernel parameter for the whole dataset is not optimal. The fault cannot be caught in the control charts. Optimising a kernel parameter for each of the variables proved to be a solution, but this is only feasible with an efficient optimiser such as the KF, as it is computationally complex to evaluate numerous parameters through grid-search. There are no instances in literature where individual kernels have been fitted for each of the variables in K-MSPC. 
    \item II. \textit{The proposed method is fast.} Even when optimising individual kernel parameters for each variable, the convergence was achieved fast, in a few hundreds of iterations and a maximum of four minutes, for a computer with the specifications mentioned in \textbf{Appendix B}.
    \item III. \textit{The model is more interpretable.} By fitting individual parameters for each of the process variables, there are insights into the amount of variation utilised by the model form each individual variable. This can be an indicator of variable importance in fault discrimination and can aid in fault diagnostics.
    \item IV. \textit{There is no need to know the optimal kernel function in advance}. If there is no prior knowledge on the optimal kernel, a combination of kernels can be utilised. In most of the existing literature, only the Gaussian kernel function is utilised, which is only applicable if data is smooth. If data contains outliers, a Cauchy kernel is more suitable. If there is a need for a rougher kernel, a low-$\nu$ Mat{\'e}rn kernel can be utilised.
\end{itemize}

The main limitation of the method is shown when compared to the traditional MSPC, where only the normal functioning data was needed to calibrate and train the MSPC control charts. 

\section{Conclusion}\label{ssec:conclusion}

The aim of the present study was to propose an optimisation technique for Kernelized Multivariate Statistical Process Control (K-MSPC). The proposed method is based on learning a Kernel-Principal Component Regression (K-PCR) model that discriminates between the normal-operating and faulty-operating process. The recently-developed technique Kernel Flows (KF) has been utilised for the task. The optimised kernel parameters are then utilised in calibrating the K-MSPC control charts, resulting in a better fault identification.

In Kernel MSPC, previous studies have proposed fitting a single kernel parameter for the whole process dataset. In the present study, we have demonstrated that fitting individual kernel parameters for each variable improves the fault detection and adds a layer of interpretability to the model. The method has been tested with the benchmark Tennessee Eastman Process dataset and has shown success in detecting even the most difficult faults, including the ones that have not been captured by the original study of Russell \textit{et al.}, 2000. \cite{russell2000fault}.

\newpage
\appendix
\section{Process variables} \label{app:variables}

\begin{table}[H]
\centering
\caption{Process variables.}\label{tab:processdata}%
\begin{tabular}{|c|c|c|c|}
\hline
\textbf{Legend} & \textbf{Description} & \textbf{Legend} &  \textbf{Description} \\
\hline
x01 & Time (h) & x28& Component E to Reactor (mol \%) \\
x02 & A Feed (kscmh) & x29 & Component F to Reactor (mol \%) \\
x03 & D Feed (kg/h) & x30 & Component A in Purge (mol \%) \\
x04 & E Feed (kg/h) & x31 & Component B in Purge (mol \%) \\
x05 & A and C Feed (kscmh) & x32 & Component C in Purge (mol \%) \\
x06 & Recycle Flow (kscmh) & x33 & Component D in Purge (mol \%) \\
x07 & Reactor Feed Rate (kscmh) & x34 & Component E in Purge (mol \%) \\
x08 & Reactor Pressure (kPa gauge) & x35 & Component F in Purge (mol \%) \\
x09 & Reactor Level (\%) & x36 & Component G in Purge (mol \%) \\
x10 & Reactor Temperature (°C) & x37 & Component H in Purge (mol \%) \\
x11 & Purge Rate (kscmh) & x38 & Component D in Product (mol \%) \\
x12 & Product Sep Temp (°C) & x39 & Component E in Product (mol \%) \\
x13 & Product Sep Level (\%) & x40 & Component F in Product (mol \%) \\
x14 & Product Sep Pressure (kPa gauge) & x41 & Component G in Product (mol \%) \\
x15 & Product Sep Underflow (\(m^3/h\)) & x42 & Component H in Product (mol \%) \\
x16 & Stripper Level (\%) & x43 & D feed (\%) \\
x17 & Stripper Pressure (kPa gauge) & x44 & E Feed (\%) \\
x18 & Stripper Underflow (\(m^3/h\)) & x45 & A Feed (\%) \\
x19 & Stripper Temp (°C) & x46 & A and C Feed (\%) \\
x20 & Stripper Steam Flow (kg/h) & x47 & Compressor recycle valve (\%) \\
x21 & Compressor Work (kW) & x48 & Purge valve (\%) \\
x22 & Reactor Coolant Temp (°C) & x49 & Separator liquid flow (\%) \\
x23 & Separator Coolant Temp (°C) & x50 & Stripper liquid flow (\%) \\
x24 & Component A to Reactor (mol \%) & x51 & Stripper steam valve (\%) \\
x25 & Component B to Reactor (mol \%) & x52 & Reactor Coolant (\%) \\
x26 & Component C to Reactor (mol \%) &  &    \\
x27 & Component D to Reactor (mol \%) &  &    \\
\hline
\end{tabular}
\end{table}

\section{ Computational resources}\label{ssec:AppendixB}

The simulations have been ran on a \textbf{MacBook Pro} with an Apple M2 chip,  16 GB/RAM. The scripts were coded in the Julia programming language and ran in Visual Studio Code. 

\section{Optimising both the scalar and the kernel width}\label{ssec:appendixC}

\begin{figure}[H]
    \centering
    \begin{subfigure}[b]{0.9\linewidth}
        \includegraphics[width=\linewidth]{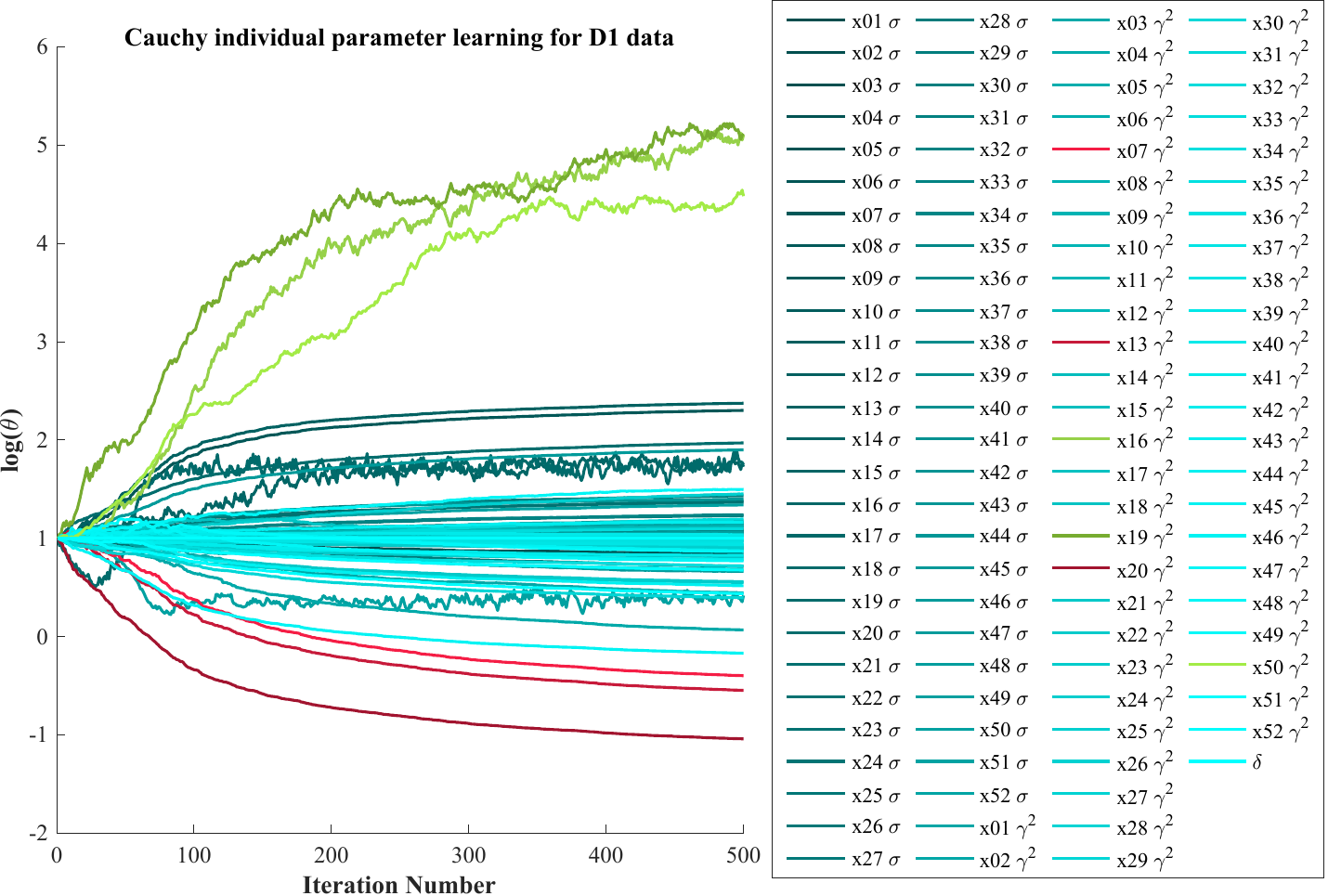}
        \caption{}
        \label{fig:paramLearningAppend}
    \end{subfigure}
    \hfill
    \begin{subfigure}[b]{0.7\linewidth}
        \centering
        \includegraphics[width=\linewidth]{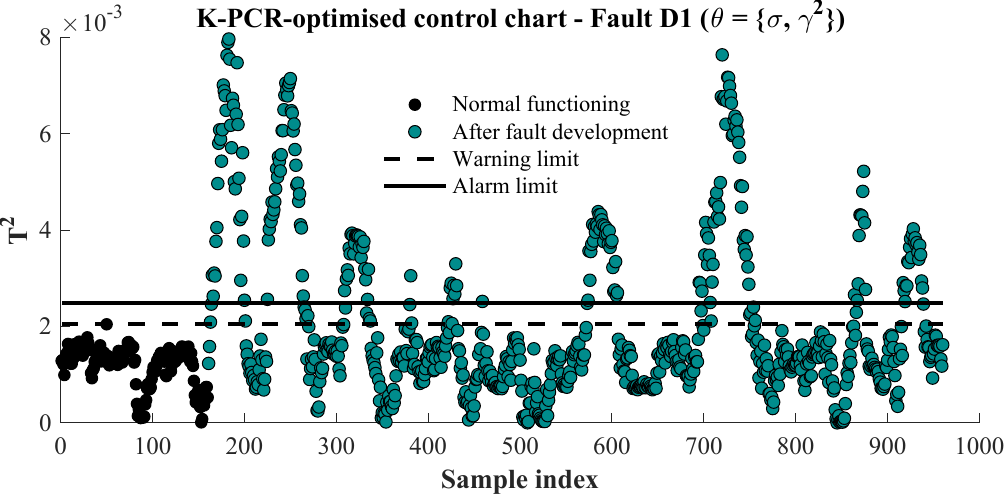}
        \caption{}
        \label{fig:controlChartAppendix}
    \end{subfigure}
    \caption{Individual parameter learning for a Cauchy kernel with $\boldsymbol{\theta} = \{\boldsymbol{\sigma}, \boldsymbol{\gamma}\}$ (a), and the resulting control chart (b).}
    \label{fig:withScalar}
\end{figure}

While results of the \textbf{D1} optimisation are presented in Subsection~\ref{ssec:KFMSPC} for a Cauchy kernel fitted individually for each variable without a scalar, the present section aims to evaluate the scenario where a scalar $\gamma^2$ is added for each of the individual-variable kernels (Fig.~\ref{fig:withScalar}).  

As observed in Fig.~\ref{fig:paramLearningAppend},
the scalar parameters ($\gamma^2$) of variables \textbf{x19}, \textbf{x16}, and \textbf{x50} converged in the highest values. They represent the stripper temperature, the stripper level and the stripper liquid flow. While there is no measured variable for the property that is driving the fault (the D feed temperature), the effects of this fault are indirectly seen in the abnormal functioning of the stripper. Amongst the variables that got the lowest scalar values (\textbf{x20}, \textbf{x13}, and \textbf{x07}), there is the stripper steam flow, the product sep. level, and the reactor feed rate.

The cost of fitting 52 more parameters comes with a slight improvement in the control chart, which has no false warnings and a good capture rate for the fault, as seen in \ref{fig:controlChartAppendix}.

\section{Comparison with other optimisation methods}

The present section presents results for the case where the kernel parameters are optimized with the \textbf{E1} and \textbf{D1} faults, and new faults are projected onto the control charts (\textbf{E2} and \textbf{D2}). The Kernel-Flows results for these are presented in Subsection~\ref{ssec:KFMSPC}, and this extension aims to present results with other benchmark optimisers for the same scenarios.

\begin{figure}[H]
    \centering
    \begin{subfigure}[b]{0.69\linewidth}
        \includegraphics[width=\linewidth]{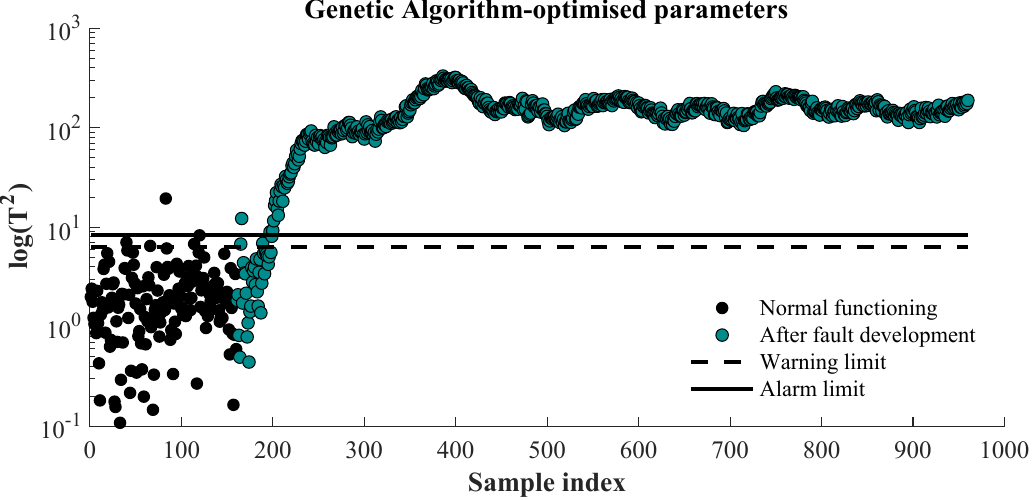}
        \caption{\textbf{E2} fault}
        \label{fig:E2t2GA}
    \end{subfigure}
    \centering
    \begin{subfigure}[b]{0.69\linewidth}
        \includegraphics[width=\linewidth]{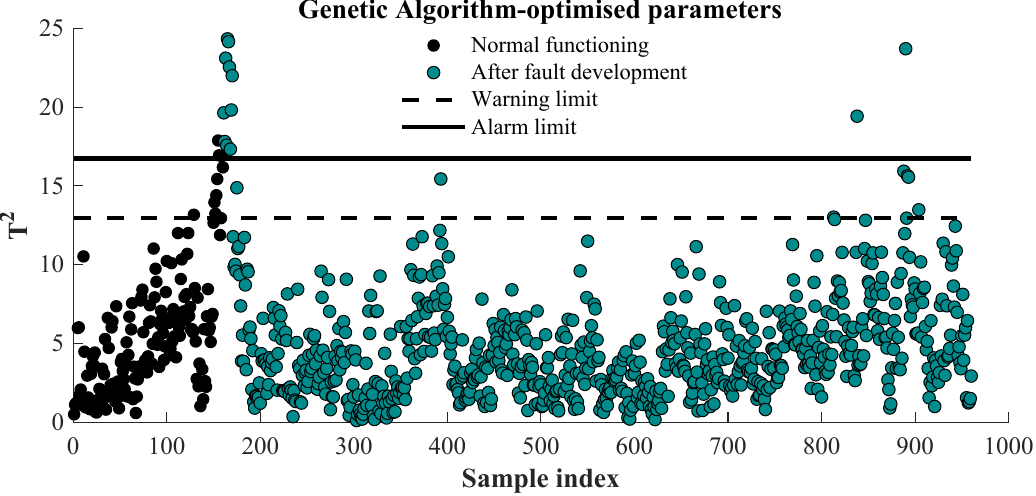}
        \caption{\textbf{D2} fault}
        \label{fig:D2t2GA}
    \end{subfigure}
    \caption{The control charts resulted by parameter optimisation with Genetic Algorithm, following the procedure and cost function from Jia \textit{et al.}, \cite{jia2012optimization}. The optimal kernel function is the Gaussian kernel.}.
    \label{fig:plotsGA}
\end{figure}

\begin{figure}[H]
    \centering
    \begin{subfigure}[b]{0.69\linewidth}
        \includegraphics[width=\linewidth]{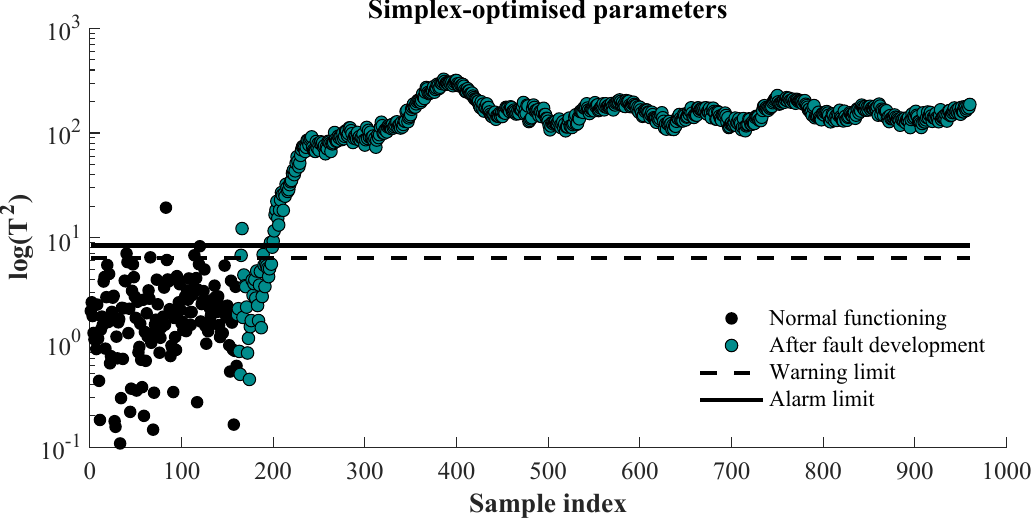}
        \caption{}
        \label{fig:e2t2sIM}
    \end{subfigure}
    \begin{subfigure}[b]{0.69\linewidth}
        \includegraphics[width=\linewidth]{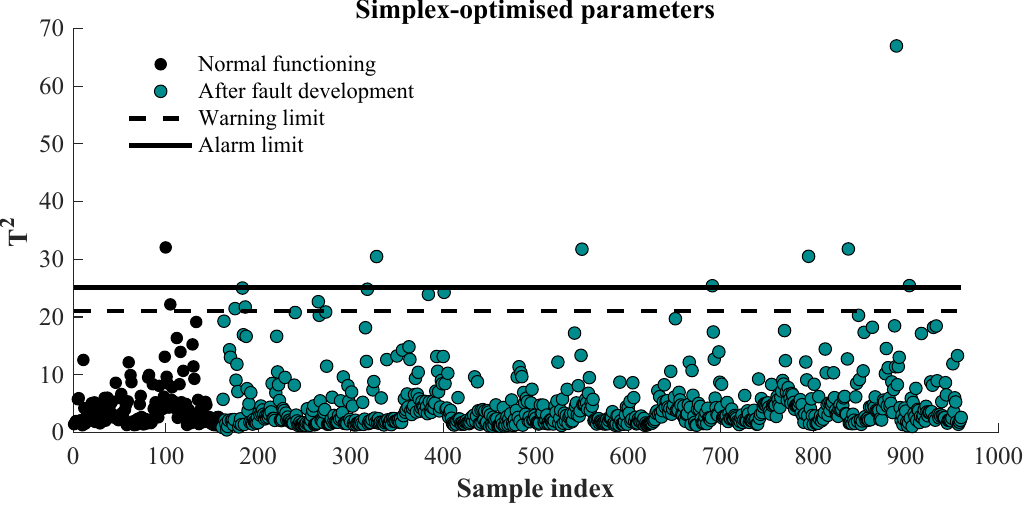}
        \caption{}
        \label{fig:d2t2SIM}
    \end{subfigure}
    \caption{The control charts resulted by parameter optimisation with Nedler-Mead simplex algorithm.}
    \label{fig:plotsSIM}
\end{figure}

For the easy-to-detect fault \textbf{E1}, both the Genetic Algorithm (Fig.~\ref{fig:E2t2GA}) and the Nedler-Mead Simplex optimizers (Fig.~\ref{fig:e2t2sIM}) have converged in similar kernel parameters, which make the control charts almost identical. 

For the difficult-to-detect case (\textbf{D2}), however, difference between the two benchmark optimizers increases. In the case of the Genetic Algorithm optimization presented in Fig.~\ref{fig:D2t2GA}, the MSPC captures a few measurements after the fault development, and then the process appears to be in-control. The Nedler-Mead simplex optimizer is not able to fit parameters so that the fault is captured, as observed in \ref{fig:d2t2SIM}. The KF-optimized K-PCR method introduced in the present paper is able to outperform both of the benchmark optimizers.

\section*{Acknowledgements}
We acknowledge funding from the Research Council of Finland for the Centre of Excellence in Inverse Modelling and Imaging 2018--2025 (decision number 353095) and for the Flagship of Advanced Mathematics for Sensing, Imaging, and Modelling 2024--2031 (decision number 359183). VJ was supported through the Higher Education for Economic Transformation (HEET) program, funded by the World Bank through the Government of Tanzania.

\section*{Autor contributions}
The co-author contributions of the paper are:

\begin{itemize}
    \item \text{Zina-Sabrina Duma}: Conceptualization; Data curation; Formal analysis; Investigation; Methodology; Software; Validation; Visualization; Roles/Writing - original draft; and Writing - review \& editing;
    \item \text{Victoria Jorry}: Conceptualization; Data curation; Formal analysis; Investigation; Methodology; Software; Validation; Visualization; Roles/Writing - original draft; and Writing - review \& editing;
    \item \text{Tuomas Sihvonen}:  Conceptualization; Formal analysis; Investigation; Validation; Visualization; Roles/Writing - original draft; and Writing - review \& editing;
    \item \text{Satu-Pia Reinikainen}: Conceptualization; Formal analysis; Funding acquisition; Methodology; Project administration; Resources; Supervision; Validation;
    \item \text{Lassi Roininen}: Conceptualization; Formal analysis; Funding acquisition; Methodology; Project administration; Resources; Supervision; Validation;
\end{itemize}  

\section*{Data availability}

The data and the developed toolbox has been made publicly available in GitHub at \\ https://github.com/sab-in-science/k-mspc.

\bibliographystyle{elsarticle-num} 
\bibliography{cas-refs}

\end{document}